\documentstyle[epsfig,12pt]{article}
\newlength{\dinwidth}
\newlength{\dinmargin}
\newcommand{\ba}{\begin{array}}
\newcommand{\ea}{\end{array}}
\newcommand{\be}{\begin{equation}}
\newcommand{\ee}{\end{equation}}
\newcommand{\bea}{\begin{eqnarray}}
\newcommand{\eea}{\end{eqnarray}}

\def\bee{\begin{eqnarray}}
\def\eee{\end{eqnarray}}
\def\be{\begin{equation}}
\def\ee{\end{equation}}

\newcommand{\beas}{\begin{eqnarray*}}
\newcommand{\eeas}{\end{eqnarray*}}

\font\cmss = cmss12

\def\integer{{\rlap{\cmss Z} \hskip 1.8pt \hbox{\cmss Z}}}
\def\laplace{{\kern1pt\vbox{\hrule height 1.2pt\hbox{\vrule width 1.2pt\hskip
  3pt\vbox{\vskip 6pt}\hskip 3pt\vrule width 0.6pt}\hrule height 0.6pt}
  \kern1pt}}
\def\scriptlap{{\kern1pt\vbox{\hrule height 0.8pt\hbox{\vrule width 0.8pt
  \hskip2pt\vbox{\vskip 4pt}\hskip 2pt\vrule width 0.4pt}\hrule height 0.4pt}
  \kern1pt}}

\def\roughly#1{\raise.3ex\hbox{$#1$\kern-.75em\lower1ex\hbox{$\sim$}}}

\begin{document}
\thispagestyle{empty}
\addtocounter{page}{-1}
\begin{flushright}
SNUTP 97-156\\
{\tt hep-th/9711202}\\
\end{flushright}
\vspace*{1.3cm}
\centerline{\Large \bf BPS Dynamics of Triple $(p,q)$ String Junction
\footnote{
Work supported in part by the NSF-KOSEF Bilateral Grant, 
KOSEF SRC-Program, Ministry of Education Grant BSRI 97-2418, 
and the Korea Foundation for Advanced Studies Faculty Fellowship.}}
\vspace*{1.2cm} \centerline{\large\bf Soo-Jong Rey and Jung-Tay Yee}
\vspace*{0.8cm}
\centerline{\large\it Physics Department} 
\vskip0.3cm
\centerline{\large \it Seoul National University, Seoul 151-742 KOREA}
\vspace*{0.6cm}
\centerline{\large\tt jungtay@phya.snu.ac.kr, sjrey@gravity.snu.ac.kr}
\vspace*{1.5cm}
\centerline{\large\bf abstract}
\vskip0.5cm
We study dynamics of triple junction of $(p,q)$ strings in Type IIB string 
theory. We probe tension and mass density of $(p,q)$ strings by studying
harmonic fluctuations of the triple junction. We show that they agree perfectly 
with BPS formula {\sl provided} suitable geometric interpretation of the 
junction is given. We provide a precise statement of BPS limit and 
force-balance property. At weak coupling and sufficiently dense limit, we 
argue that $(p,q)$-string embedded in string network is a `wiggly string', 
whose low-energy dynamics can be described via renormalization group evolved, 
smooth effective non-relativistic string. We also suggest the possibility 
that, upon Type IIB strings are promoted to M-theory membrane, there can exist 
`evanescent' bound-states at triple junction in the continuum. 
\vspace*{1.1cm}

\setlength{\baselineskip}{18pt}
\setlength{\parskip}{12pt}

\newpage
\vskip 1.0 cm
{\tt Triple String Junction}: 
Among Dirichlet branes, the quantum solitons in string theory, 
Type IIB $(p,q)$ string is of particular interest in that it is 
the simplest {\sl non-threshold} bound-state consisting of $p$ fundamental
strings (F-string) and $q$ Dirichlet strings (D-string).
A $(p,q)$-string is BPS saturated, hence, string tension equals to mass
density and is given by~\cite{schwarz}
\be
T_{(p,q)} = T \sqrt{p^2 + {q^2 \over g^2_{\rm IIB}}},
\label{bpstension}
\ee
where $T = T_{(1,0)} \equiv 1 / 2 \pi \alpha'$ is the F-string tension and $g_{\rm IIB}$
denotes Type IIB string coupling parameter. Indeed, for relatively prime 
integers $p$ and $q$, the entire string bound-states form an orbit of 
$SL(2,\integer)$ S-duality in Type IIB string theory~\cite{schwarz,witten}.

Because of non-threshold nature of the bound-state, formation of
$(p,q)$-string entails intriguing interplay between gauge field on the D-string
worldsheet and F-string charge density~\cite{witten}. To release latent 
binding energy into local recoil~\cite{leerey}, the F-string charge had better 
spread over the D-string worldsheet. This is made possible via Cremmer-Scherk 
coupling~\cite{cremmerscherk} that allows transmutation of the fused F-string 
charge to electric flux of D-string worldsheet gauge field.  

Consider the fusion process of a F-string onto a D-string. If only part of 
the F-string is bound to the D-string and is transmuted to worldsheet gauge 
field, the resulting geometry is nothing but a triple junction of $(p,0)$ 
F-string, $(0,q)$ D-string and $(p,q)$-string. The configuration is depicted
schematically in Fig. 1(a). At each junction F- and D-string charges are
conserved separately~\cite{yankielowicz}:
\be
\sum_{a=1}^3 p_a = 0 \quad : \quad \quad \sum_{a=1}^3 q_a = 0.
\label{conserv}
\ee
Generically the F-string will not just stop at the configuration of Fig.1(a)
but continue fusion process until they form a bound-state of $(p,q)$-string.
However, if configuration is such that the tension is balanced
\be
\sum_{a=1}^3 T_{(p_a, q_a)} = 0 \,\, ,
\label{balance}
\ee
(where the string tension is treated as a complex quantity) the triple 
junction can be stabilized and become a BPS saturated configuration.
It was previously conjectured~\cite{schwarz2}
that a triple string junction satisfying Eqs.(~\ref{conserv},\ref{balance})
is BPS saturated. Recently, in the linearized
approximation, this conjecture has been proven both in
worldsheet~\cite{dasguptamukhi} and spacetime~\cite{sen} approaches.
The former approach was inspired by recent new understanding of deformed
branes~\cite{callanmaldacena,gibbons,larus,hashimoto}.

In this Letter, we investigate {\sl BPS dynamics} of the triple string 
junction. We do so by studying dynamics of harmonic fluctuation and extract
information relevant to BPS condition of the triple junction. For a single
string, in physical gauge, harmonic fluctuation is governed by an action
\be
{\cal S} = \int d t d \sigma \, \left[
{1 \over 2} \rho \left( \partial_t {\bf X} \right)^2   
- {1 \over 2} T \left( \partial_\sigma {\bf X} \right)^2 \right].
\label{generic}
\ee
Here, $\rho$ and $T$ denote {\sl inertial} mass density and tension of the 
string, that are in general functions of $(t, \sigma)$. Dynamics of the string is completely characterized only when the action
Eq.(\ref{generic}) is supplemented by an `equation of state' $T = T(\rho)$. 
For example, propagation velocities of transverse and longitudinal oscillations 
are given by 
$v_{\rm T}^2 \equiv T(\rho) / \rho \quad; \quad
v_{\rm L}^2  \equiv - d T (\rho) /  d \rho \, \, .
$
For a fundamental BPS string, longitudinal oscillation is gauge 
redundant and $T(\rho) = c^2 \rho$ is $(t, \sigma)$-independent constant.
However, this is no longer true for
coarse-grained effective description of `wiggly $(p,q)$-string', on
which we will return momentarily.
Is it possible to probe the string equation of state $T = T(\rho)$,
dynamics for each string prong, and for the triple junction as a whole?
We now answer these questions affirmatively
and gain new understanding of {\sl BPS dynamics} of triple string junction
that are visible only in full-fledged Dirac-Born-Infeld (DBI) analysis.


\begin{figure}[t]
   \vspace{0cm}
   \epsfysize=7cm
   \epsfxsize=13cm
   \centerline{\epsffile{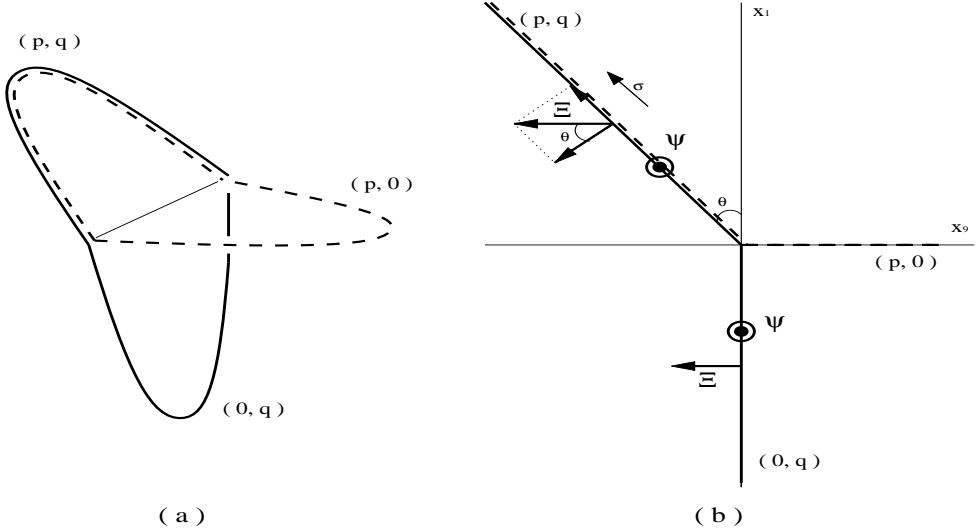}}
\caption{\label{fig1} BPS geometry of triple $(p,q)$-string junction}
\end{figure}

{\tt Dirac-Born-Infeld Analysis:}
Consider a $(0,q)$ D-string aligned initially along $x_1$ direction and 
a $(p,0)$ F-string along $x_9$ direction impinging on D-string at $x_1 
= x_9 = 0$. See Fig. 1(b). In the static gauge $X^0 = t, \, X^1 = x_1$,
worldsheet dynamics of the $(0,q)$ D-string is described by the following
abelian part~\cite{larus} of the DBI Lagrangian:
\be
L_{\rm DBI} = - {q \over g_{\rm IIB}} T
\int d x_1 \, \sqrt{ \det(\eta_{ab} + F_{ab} + \nabla_a {\bf X} \cdot 
\nabla_b {\bf X})} .
\ee 
We have chosen a static gauge $X^0 = t, \, \, X^1 = x_1, \,\, A_1 = 0$,
and have denoted the transverse collective coordinates as ${\bf X}$ and 
the worldsheet gauge field as $F_{ab}$.
Assuming that classical configuration excites ${\bf X}$ along the plane
spanned by injected F-string and D-string (9-direction) and is static,
the total energy functional of D-string is given by
\be
H_{\rm DBI} = {q \over g_{\rm IIB}} T \int d x_1 \,
{1 + \left( \nabla_1 X^9 \right)^2 \over
\sqrt{ 1 + \left(\nabla_1 X^9 \right)^2 - \left( \nabla_1 A_0 \right)^2}
} \,\,\ .
\label{dbienergy}
\ee
Extremum of the energy functional is when $A_0$ and $X^9$ satisfy
\bee
\sqrt{ a} \, \nabla_1 A_0 &=& \nabla_1 X^9
\nonumber \\
\nabla_1 \cdot \left( {\nabla_1 A_0 
\over \sqrt{1 + (\nabla_1 X^9)^2 - (\nabla_1 A)^2}} \right) &=& 
g_{\rm IIB} {p \over q} \, \delta (x_1)
\eee
where $a$ is a constant parameter ($0 < a < \infty$). In the last equation,
we have used the fact that $p$ units of F-string charges are distributed
among $q$-multiples of D-string. Comparison with
previous results~\cite{callanmaldacena,gibbons,hashimoto}
shows that $a = 1$ corresponds to
the BPS limit. The equations are solved straightforwardly to yield
an exact solution:
\bee
X^9 (x_1) = \sqrt{a} A_0 (x_1) 
&=& - \left( \tan {\cal \theta} \right) \, x_1 \,\, \quad \quad \quad (x > 0) 
\nonumber \\
&=& \quad \quad 0 \quad \quad \quad \quad \quad \quad (x < 0)
\label{background}
\eee
where
\be
\tan \theta \equiv {X^9 \over x_1} = 
\sqrt { \ba{c}  a \\ \ea  \over \left[ 
 (1 - a) + {1 \over g_{\rm IIB}^2 } \left(q / p\right)^2 \right]  
} .
\label{tan}
\ee
The solution has the following simple geometric interpretation. 
At the impinging point of the F-string, the initially straight D-string is
bent rigidly by angle $\theta$ to the negative $x_9$ direction.
We emphasize that the angle $\theta$ is determined solely by
the $(p,q)$ charges (at fixed $a, g_{\rm IIB}$ values), not by a requirement
of tension balance. Note also that the bending angle $\theta$
increases {\sl monotonically} with increasing the parameter $a$.
Away from the triple junction location, the D- and F-string prongs
at $x_1 < 0$ and $x_9 > 0$ are nothing but BPS saturated single string states
with $T(\rho) = \rho$ equals to $(q /g_{\rm IIB}) T $ and $p T$ respectively. 
What about the $(p,q)$ string prong, now bent to the second quadrangle?
We now show that, for {\sl all} values of $a$, the $(p,q)$-string prong itself
is also BPS saturated.
To show this, we evaluate {\sl static} mass density of the $(p,q)$ string from
Eq.(\ref{dbienergy}):
\be
E_{(p,q) \,\, \rm string}
 = \int_0^\infty d x_1 \, {q \over g_{\rm IIB}} T \, 
{1 + \tan^2 \theta \over \sqrt{1 + {a - 1 \over a} \tan^2 \theta} } \, \, .
\label{pqenergy}
\ee
One expects the integrand to represent mass density of the $(p,q)$-string prong.
However, it looks nothing like the BPS formula for any value of $a$, even
including the expected BPS limit~\cite{callanmaldacena,gibbons,hashimoto} $a=1$!
This puzzle is resolved neatly by noting that the $(p,q)$-string prong has
now been bent rigidly by angle $\theta$ relative to the $x_1$ axis.
Therefore, along
$(p,q)$-string prong, we introduce a proper worldsheet coordinate
$\sigma$ and measure the string mass density per unit 
$\sigma$-length. From elementary geometry,
\be
\sigma = {x_1 \over \cos \theta}
\quad (-\infty < \sigma < +\infty); \quad \quad {1 \over \cos \theta} = 
\sqrt{1 + {a \over 1 - a + {1 \over g_{\rm IIB}^2} \left(q / p\right)^2}
} \,\, .
\label{newcoord}
\ee
After taking this simple geometric consideration into account to
Eq.(\ref{pqenergy}) we identify proper {\sl static} mass density
$\rho_{(p,q)}$ of the $(p,q)$-string prong:
\be
E_{(p,q) \,\, \rm string}
= \int_0^\infty ds \, \rho_{(p,q)} \quad : \quad \quad
\rho_{(p,q)} \equiv T \, \sqrt {p^2 + {q^2 \over g_{\rm IIB}^2} }.
\label{pqdensity}
\ee
As claimed, for {\sl all} values of parameter $a$, we have shown that the static
mass density of the $(p,q)$-string prong of the triple junction is
BPS saturated to the tension Eq.(\ref{bpstension}):
$\rho_{(p,q)} = T_{(p,q)}$.

What is then special to the proclaimed BPS limit
$a=1$~\cite{callanmaldacena,gibbons,hashimoto}? We now show that,
even though each string prong is always BPS saturated, it is only when
$a=1$ the string tensions sum to zero so that the triple string junction stays
in equilibrium. For an arbitrary $a$, ratios of
vector components of string tensions along $x_1, x_9$ directions are
\bee
x_1\quad: \quad \quad T_{(p,q)} \cos \theta / T_{(0,q)}
&=& - \sqrt{ 1 + (1 - a) g_{\rm IIB}^2 (p/q)^2}
\nonumber \\
x_9 \quad: \quad \quad T_{(p,q)} \sin \theta / T_{(p,0)}
&=& - \sqrt{a} \quad.
\eee
It is clear that only when $a=1$ vector sum of three string tensions vanish
identically.
If $a > 1$, net force is nonvanishing and acts on the triple string
junction to the direction of third quadrant.
As the triple string junction responds adiabatically to the force,
the angle $\theta$ decreases monotonically until it reaches the BPS value
$\theta_{\rm BPS} = \tan^{-1}(p/q) g_{\rm IIB}$.
Likewise, if $a < 1$, net force acts to the first quadrant so that
the triple string junction moves in the direction of increasing $\theta$
to the BPS value. We thus conclude that any triple string junction with
$a \ne 1$ relaxes always to $a = 1$ configuration. Strictly speaking, the
assumption that triple string junction is a {\sl static}
configuration is not valid for $a \ne 1$.

\tt Harmonic Fluctuation: \rm 
Another quantities of physical interest are {\sl inertial} mass density and tension of the
triple string as defined via Eq.~(\ref{generic}).
We now probe these quantities by studying {\sl harmonic fluctuations} of the
D-string, part of which has now bent into $(p,q)$-string prong. In this
approach,
F-string serves only as a static background source of electric charge on the
D-string worldsheet. Let us decompose fluctuation of ${\bf X}^i$ around
the background Eq.(\ref{background})
into a fluctuation $\Xi$ parallel to 9-direction and perpendicular
to 1-direction and a fluctuation $\Psi$ orthogonal to 1- and 9-directions
(in- and out-of-plane fluctuations in Fig. 1(b)).
The harmonic dynamics of D-string is governed by quadratic
expansion of the DBI Lagrangian:
\bee
L_{\rm DBI}^{(2)}  =
\int_{-\infty}^{+\infty} d x_1 \, {T \over 2} \frac{q}{g_{\rm IIB}} 
\frac{1}{ \sqrt{1+ B^2 - E^2} }
\Big[ &+& {1 + B^2 \over 1 + B^2 - E^2 } \, \left( {\cal F}_{01} \right)^2 
-  2 {E B \over 1 + B^2 - E^2 } \, {\cal F}_{01} \cdot \nabla_1 \Xi  
\nonumber \\ 
 &+ & \left( \nabla_0 \Xi \right)^2 - { 1 - E^2 \over 1 + B^2 - E^2 } \,
\left( \nabla_1 \Xi \right)^2
\nonumber \\
&+&(1 + B^2) \left( \nabla_0 \Psi \right)^2
- \left(\nabla_1 \Psi \right)^2 \Big]
\eee
Here, background fields are abbreviated as $ B \equiv \nabla_1 X^9$ and 
$ E \equiv - \nabla_1 A_0$. Fluctuation of the worldsheet gauge field 
can be integrated out exactly. This yields:
\be
 L_{\rm DBI}^{(2)} =  \int_{-\infty}^{+\infty} \!\! dx_1 \,
 {T \over 2} {q  \over g_{\rm IIB} } {1 \over \sqrt{1+ B^2  - E^2} } [
\left( \nabla_0 \Xi \right)^2  - {1 \over \left( 1 + B^2 \right)} \,
\left( \nabla_1 \Xi \right)^2 
\, + \,
\left(1 + B^2 \right)\, \left(\nabla_0 \Psi \right)^2
-\left( \nabla_1 \Psi \right)^2
\Big] \, .
\ee
Comparing this with Eq.(\ref{generic}) one might be tempted to conclude that 
the {\sl inertial} mass density and tension 
differ from the {\sl static} ones Eq.~(\ref{pqdensity})
and are also sensitive to the polarization directions, $\Xi$ or $\Psi$.
We now show that this is {\sl not} the case.
Key observation is again associated with
proper geometric interpretation. As for the identification of 
static mass density, we should measure fluctuation with respect to the $\sigma$
coordinate introduced in Eq.~(\ref{newcoord}). Furthermore, for in-plane
fluctuation $\Xi$ along $x_9$ direction,
component parallel to the $(p,q)$-string prong should
be interpreted as a gauge redundant longitudinal mode, hence, only component
perpendicular to the string is physical. 
Geometrically, these considerations amount to change of variable
$x \rightarrow \sigma \cos \theta$ and projection $\Xi
\rightarrow \Xi_\perp / \cos \theta$.
Thus, we identify proper DBI Lagrangian describing harmonic fluctuation as:
\begin{eqnarray}
 L_{\rm DBI}^{(2)}  =  \int_{-\infty}^{+\infty} d \sigma \, {T \over 2} 
{q \over g_{\rm IIB}} 
        \sqrt{ \frac{ 1+ a E^2 }{ 1+ ( a-1 ) E^2 }}
        \Big[ \,\,\, (\nabla_t {\Psi})^2 - (\nabla_\sigma \Psi)^2 \,\,
+ (\nabla_t {\Xi}_\bot)^2 - (\nabla_\sigma {\Xi_\bot})^2 \,\,\, \Big] \,\, ,
\label{final}
\end{eqnarray}
where $ 1/E^2 = a \cot^2 \theta = (1-a) + (q^2/p^2)/ g^2_{\rm IIB}$.
Thus, for {\sl all} values of $a$, 
the {\sl inertial}
mass density and the tension of the fluctiation per unit length of
the $(p,q)$-string prong are equal to
$ \frac{q}{g_{\rm IIB}} T \sqrt{ \frac{ 1+ a E^2 }{ 1+ ( a-1 ) E^2 }} =
T \sqrt{ \frac{q^2}{g^2_{\rm IIB}} + p^2 } $.
They are exactly the same as the {\sl static} mass density and tension
Eqs.(~\ref{pqdensity},\ref{bpstension}).

It is instructive to repeat the above analysis for a $(0,q)$
Dirichlet $n$-brane attached by a $(p,0)$
F-string~\cite{callanmaldacena,gibbons,larus,hashimoto}.
On the world-volume of D-brane, the configuration is described by
spherically symmetric background ${\bf E} = {\hat r} \nabla_r A_0$
and ${\bf B} = {\hat r} \nabla_r X^9$ satisfying ${\bf B}
= {\sqrt a} {\bf E}$ and
\be
{1 \over {\bf E}^2} = (1 - a) + \left( { (q/p) r^{n-1} \over (n-2) c_n}
\right)^2 \,\, : \quad
c_n = {(2 \pi / T)^{(n-1)/2} \over (n-2) \Omega_{n-1} } \, g_{\rm IIB}.
\ee
For simplicity, we restrict fluctuations around the above configuration
to S-wave partial wave modes only. After integrating out world-volume gauge
field, the DBI Lagrangian of harmonic fluctuation is given by:
\be
L_{\rm DBI} = {T^{(n)}  \over 2} { q \over g_{\rm IIB}}
\int d^n r \, 
{1 \over \sqrt { 1 + {\bf B}^2 - {\bf E}^2} }
\Big[ (1 + {\bf B}^2) (\nabla_0 \Xi)^2 - (\nabla_r \Xi)^2
+ (\nabla_0 \Psi)^2 - {1 \over (1 + {\bf B}^2)} (\nabla_r \Psi)^2 \Big],
\ee 
where $T^{(n)} \equiv \Omega_{n-1} (2 \pi)^{(1-n)/2} T^{(n+1)/2}$, and
$\Xi$ and $\Psi$ denote fluctuations along $x_9$-direction and perpendicular
to $x_9$ and Dirichlet $n$-brane directions respectively.
Following the same geometric reasoning as in D-string case, proper description
of fluctuations is obtained once
we make a change of variable $r \rightarrow \sigma \cos \theta$ and
orthogonal projection $\Xi \rightarrow \Xi_\perp / \cos \theta$, where
$\cos \theta = \sqrt{1 / (1 + {\bf B}^2)}$.
The DBI Lagrangian for proper fluctuation of Dirichlet n-brane is then given by
\be
L_{\rm DBI} = {T^{(n)} \over 2} {q \over g_{\rm IIB}}
\int d^n \sigma \cos^n \theta \sqrt{ 1 + a {\bf E}^2 \over
1 + (a - 1) {\bf E}^2} 
\, \Big[ (\nabla_0 \Psi)^2 - (\nabla_\sigma \Psi)^2
\, + \, (\nabla_0 \Xi_\perp)^2 - (\nabla_\sigma \Xi_\perp)^2 \, \, ].
\ee
Therefore, for all possible polarization, we find that the triple Dirichlet 
$n$-brane junction is a BPS saturated configuration with equal {\sl inertial}
mass density and tension:
\be
\rho^{(n)}_{(p,q)} = T^{(n)}_{(p,q)}
= T^{(n)} \sqrt{ \left( {(2 \pi / T)^{(n-1)/2} \over \Omega_{n-1} r^{(n-1)} }
\right)^2 \, p^2 + {q^2 \over g_{\rm IIB}^2} }.
\label{nbranedensity}
\ee
A novelty not encountered for triple $(p,q)$ string is that the tension varies
continuously as one moves in from asymptotic infinity to the center, where
the F-string is impinging on the Dirichlet $n$-brane. The bending angle
$\theta$ is position-dependent and increases monotonically from zero to
maximum value,
$\theta^* = \cos^{-1} \sqrt{1/(1 - a)}$ (for $a < 1$) or $\pi / 2$ (for
$a > 1$).
Far away from the center, the BPS mass density Eq.~(\ref{nbranedensity})
approaches that of
an isolated Dirichlet $n$-brane: $(q / g_{\rm IIB}) T^{(n)}$. Near the center,
the BPS mass density per unit $n$-dimensional volume
Eq.(\ref{nbranedensity}) diverges.
However, mass density and tension measured per unit $\sigma$-length is
finite. For example, for $a < 1$,  $r^{(n-1)} T^{(n)}_{(p,q)} \rightarrow
p \, T$, the mass density of $(0,p)$ F-string. It
is straightforward to recognize that
tension is balanced at {\sl every point} on the Dirichlet $n$-brane only when
$a=1$ but not for $a \ne 1$. Hence, we interpret this as an indication that
deformed Dirichlet $n$-brane with $a \ne 1$ always relax to stable $a=1$
configuration. Details will be reported elsewhere.

{\tt Wiggly (p,q) String}:
Using the triple string junction as a buliding block, one can construct
a network of $(p,q)$-strings. Indeed, Sen~\cite{sen} has suggested the
string network
as a novel mechanism of string compactification. Being so, 
{\sl dynamical aspect} of the string network is also of interest.
Consider, for definiteness, a {\sl dense} string network 
in weakly coupled Type IIB string theory. In this limit, any prongs carrying
D-string Ramond-Ramond charge ($q \ne 0$) will become much heavier than those 
carrying F-string charge only. Because of the D-string charge conservation,
these heavy prongs should either form a closed loop or extend to infinity.
See Fig. 2(a). What are then distinguishing characteristics, if any, of a heavy
loop in the network from a $(p,q)$-string in isolation? We now argue that
coarse-grained picture of the heavy-prong loop is a smooth, non-relativistic 
string by showing that `wiggles' present on the loop renormalize the 
microscopic equation of state $T(\rho) = \rho$ to a non-trivial 
renormalization-group fixed point  $T(\rho)
\rho = T_{(p,q)}^2$ in the infrared.

We begin with an argument that the heavy-prong loop is constantly wiggled 
by the background F-string network. Consider low-energy excitations of the 
BPS string network. One possible excitation 
is to boost prongs in each triple string junction while maintaining balance 
of the tension forces. An example involving 4 adjacent junctions is illustrated 
in Fig. 2(b).  At macroscopic scale on the heavy-prong loop, net effect of 
such gapless excitation is to put wiggles to the loop. Generically, wiggles 
of all possible sizes will be present. Hence, we will call the loops made 
out of heavy prongs as `wiggly strings' -- they are nothing but D-strings 
with small scale structures induced by dense F-string network background. 

\begin{figure}[t]
   \vspace{0cm}
   \epsfysize=6cm
   \epsfxsize=14.5cm
   \centerline{\epsffile{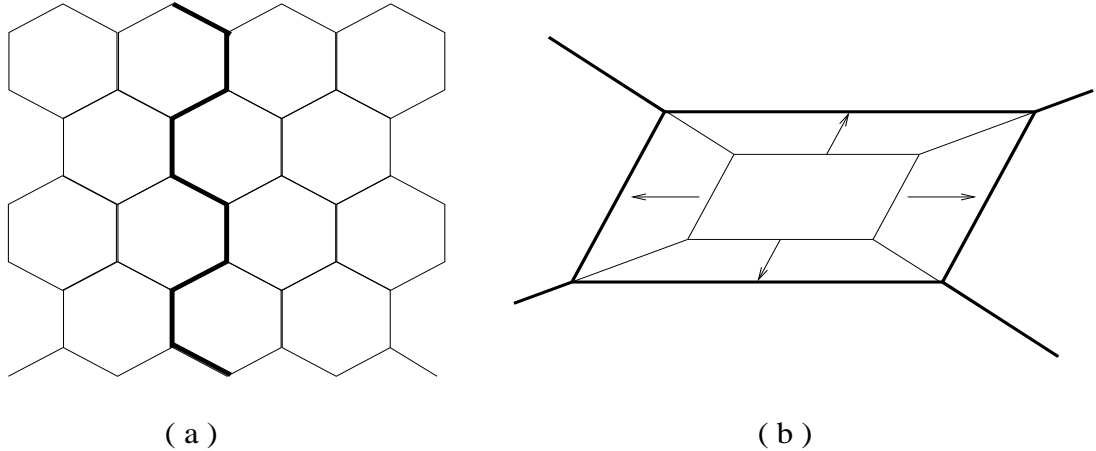}}
\caption{\label{fig2}
(a) massive $(p,q)$ string embedded in string network, (b) noise generation
via lowest-energy fluctuation.}
\end{figure}

Suppose the heavy-prong loop has wiggles on it with characteristic size
$\lambda$ but is straight otherwise. To an observer with resolution 
$\ell \gg \lambda$, the wiggly string appears to be a straight string with 
an effective mass density $\rho_{\rm eff} > \rho_{(p,q)}$ and tension 
$T_{\rm eff} < T_{(p,q)}$.
Low-energy dynamics of the wiggly string is most conveniently described
by a {\sl coarse-grained, effective smooth string} in which the 
small-scale wiggles of sizes $\lambda < \ell$ are integrated out.
Intuitively, the wiggles increase string mass density but decrease string 
tension. Hence, we expect that the microscopic equation of state 
$T (\rho) = \rho$ is unstable under coarse-graining and flows into a 
renormalization-group fixed point in the infrared. 
It is our aim to find out non-trivial infrared fixed point, if present. 
A similar question has been addressed previously~\cite{sikivie}
in the context of noisy cosmic and Nambu-Goto strings~\cite{vilenkin}.

For small amount of wiggliness, the effective parameters have the following
schematic form:
\bee
\rho_{\rm eff} &=& \rho_{(p,q)} + \langle {\bf V}^2 \rangle \, 
F (\rho_{(p,q)}, T_{(p,q)}, \cdots)
\nonumber \\
T_{\rm eff} &=& T_{(p,q)} - \langle {\bf V}^2 \rangle \, 
G (\rho_{(p,q)}, T_{(p,q)}, \cdots) .
\label{renorm}
\eee
Here,
$\langle {\bf V}^2 \rangle$ denotes the average velocity-squared 
(both transverse and longitudinal), and can be calculated from harmonic
fluctuations of the wiggly string. The $F, G$ are positive-definite 
functions of string mass density and 
tension that are determined solely by the microscopic equation of state.

At first sight, it appears that wiggly $(p,q)$-string under consideration should
be significantly different from Nambu-Goto string studied in 
Ref.~\cite{sikivie} -- for example, $(p,q)$-string has nontrivial worldsheet
gauge field excitations. However, this is {\sl not} the case. We have shown
already that, in deriving the proper DBI Lagrangian Eq.~(\ref{final}), it
was crucial not to discard the worldsheet gauge field fluctuations but 
integrate them out explicity. Harmonic fluctuation described by the resulting 
DBI Lagrangian is exactly the same as that of Nambu-Goto string. 
Since all one needs for determining the structure of Eq.~(\ref{renorm}) are 
harmonic fluctutation $\langle {\bf V}^2 \rangle$ and microscopic equation 
of state, the renormalization group equation can be derived straightforwardly
by endowing scale dependence to string mass density and tension and utilizing 
the techniques of Ref.~\cite{sikivie}. Demanding that energy-momentum 
conservation and equations of motion are satisfied for the wiggly string
with worldsheet 2-velocity $u^a(\sigma) \, (u^a u_a = + 1)$, we find 
the renormalization-group equation in Fourier mode $k$-space at the lowest
non-trivial order: 
\bee
{d \over d \ln k} \rho(k) &=& - W_T (k) \rho(k) 
 - W_L (k) \left\{ \rho (k) - T(k) \right\} + \cdots  
\nonumber \\
{d \over d \ln k} T (k) &=& - {1 \over 2} W_T (k)  
\left\{ \rho (k) + T(k) \right\} 
 + W_L (k) \left\{ \rho (k) - T (k) \right\} + \cdots,
\label{rge}
\eee
where
power-spectrum of effective transverse and longitudinal fluctuations
is given by:
\bee
W_T(k) &\equiv& k \int_0^\ell {d \sigma \over \ell} \, e^{i k \sigma} 
\Big[ \, 7 \langle \, \nabla_0 \Psi(\sigma) \cdot  \nabla_0 \Psi(0)  
\, \rangle + 
\langle \, 
\nabla_0  \Xi_\perp (\sigma) \cdot \nabla_0 \Xi_\perp (0) \,\rangle
\, \Big]
\nonumber \\
W_L (k) &\equiv& k \int_0^\ell {d \sigma \over \ell} \,
e^{i k \sigma} 
\Big[ \, \langle  \, \left(u^1(\sigma) / u^0 (\sigma) \right) \cdot \left(
u^1( 0 ) / u^0 (0) \right) \, \rangle \, \Big].
\eee
We have ignored second- or higher-order corrections on the right-hand side
of Eq.~(\ref{rge}). 
It should then become clear that, using the microscopic equation of state
$T(\rho) = \rho \equiv T_{(p,q)}$ as the ultra-violet boundary condition of
the renormalization-group equations, 
there exists a non-trivial infrared fixed point, where
\be
T(k) \rightarrow T_{\rm IR} \ll T_{(p,q)} \,\, , \quad
\rho(k) \rightarrow \rho_{\rm IR} \gg T_{(p,q)} \quad \quad
{\rm such} \,\,\,\, {\rm that}
\quad \quad
 T_{\rm IR} \cdot \rho_{\rm IR} = T_{(p,q)}^2.
\ee  
Note that, even though the microscopic $(p,q)$-string has no physical 
longitudinal excitation, effective smooth string acquires non-trivial
longitudinal dynamics via coarse-graining and has propagation velocity 
$v^2_L = - dT(\rho) / d\rho$. The infrared fixed point is characterized
by the fact that the longitudinal propagation velocity become
equal to the transverse propagation velocity, $v^2_T , v^2_L 
\rightarrow \left(T_{(p,q)} / \rho_{\rm IR} \right)^2$. We note that this is 
much less than speed of light, hence,  conclude that low-energy excitation of
wiggly $(p,q)$-string is conveniently described by a non-relativistic dynamics 
of effective smooth string.  

{\tt M-Theory Limit and Evanescent Bound-State}:
We finally discuss a possible existence of novel bound-state in the middle
of continuum excitations once the triple string junction is promoted to 
M-theory configuration. It has been
known~\cite{yankielowicz, schwarz2} that the triple string junction arises
from M-theory by starting with a "pant" configuration of membrane and 
wrapping each of the membrane prongs on different cycles of compactified
two-dimensional torus. 

\begin{figure}[t]
   \vspace{0cm}
   \epsfysize=6cm
   \epsfxsize=8cm
   \centerline{\epsffile{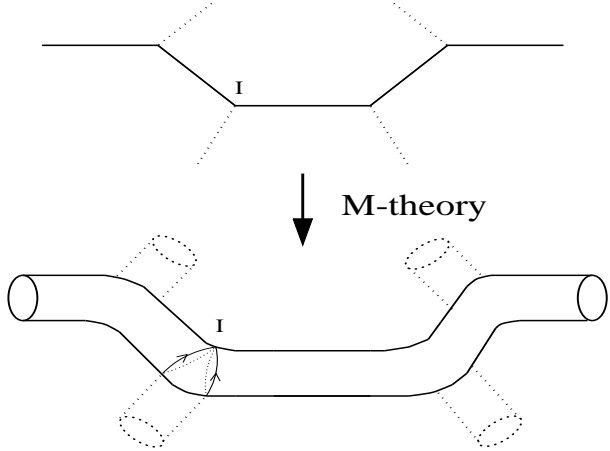}}
\caption{\label{fig3} 
Triple string junction in M-Theory Limit}
\end{figure}

Consider a D-string impinged by several F-strings. In the M-theory limit, the 
configuration look like a `twisted tube' as depicted in Fig. 3 (only the cycle
associated with D-string charge is shown explicitly). Excitations of such a 
triple membrane junction is carried by waves propagating on the surface of 
`twisted tube'. Along the direction of common ${\bf S}_1$ cycle, 
the waves satisfy periodic boundary condition. Hence, for each periodic
normal modes, propagation of low-energy waves along the tube direction is 
described by a one-dimensional Helmholtz equation.
Is there any new phenomena due to the fact that the tube is twisted instead
of being straight? We now provide an argument that suggests a positive answer 
to this question.

For definiteness, we will take the limit that the common ${\bf S}_1$ cycle
is small but non-zero. In describing the propagation of effective 
one-dimensional waves along the direction of the twisted tube (See Fig. 3),
it is more convenient to use the natural coordinate $\sigma$ introduced
earlier in Eq.(11). Crucial point is that the projection angle $\theta$ varies
every time one passes the triple membrane junction. Geometrically, it is clear 
then that the junction region (denoted as I in Fig. 3) has effectively larger 
surface of the tube than the region in between. This results in {\sl relative 
decrease} of the normal-mode frequency in the junction region compared to 
the membrane-prong or asymptotic prong regions.
More concretely, let us start from a small fluctuation equation of motion 
of the membrane in terms of physical gauge coordinates 
$x_0, x_1, x_2$ (the analogs of $x_0, x_1$ in Eq.(8)). If we make a change 
of variables to natural {\sl curved} coordinates $\sigma$ along the tube
direction,  the effective one-dimensional wave equation takes the following 
form:
\be
\left[ \partial_t^2 - \partial_\sigma^2 - V(\sigma) + (k_{2n})^2 \right]
{\bf X}^i_n (\sigma, t) = 0,
\ee
where
$k_n$ denotes the normal frequency around the common ${\bf S}_1$ direction
and ${\bf X}^i_n$ is the normal mode of membrane fluctuation. The crucial
term  $V(\sigma) = \left( \kappa (\sigma) / 2 \right)^2$ is a
potential is induced during the course of the change of variables from the 
ambient flat 
space coordinate $x_1, x_2$ to the curved one $\sigma, x_2$, 
and is expressible in terms of
extrinsic curvature $\kappa$ of the tube~\cite{boundstate,boundstate2}.
It is transparent that the induced potential can be interpreted as a local
decrease of the normal frequency $(k_{1m})^2$. If the induced potential is
smooth enough, $V(\sigma) \approx \langle V \rangle$= constant, 
effective dispersion relation locally in the neighborhood of the triple 
membrane junction is 
\be
\omega^2 =  (\tilde k_{1m})^2 + (k_{2n})^2
\quad \quad {\rm where} \quad \quad
(\tilde k_{1m})^2 \equiv (k_{1m})^2 - \langle V \rangle.
\ee
As stated, the normal frequency along the tube direction is lowered
effectively by twisting of the membrane. In particular, when $k_{1m}^2 
\rightarrow 0$, $\tilde k_{1m}^2 < 0$ and $(k_{2n} / \omega) > 1$! 
This is an indication that there may exist effectively one-dimensional 
{\sl evanescent bound-states}~\cite{ll} in the middle of the scattering 
continuum, whose wave function is localized near 
the triple membrane junction region.
Note that these excitations are non-BPS since both $\tilde k_1$ and $k_2$ 
should be nonvanishing.
We conjecture that this is not specific to the triple membrane junction
but very generic to small fluctuations of any curved membrane.
We will report further details elsewhere.


\end{document}